\documentclass[10pt,final,journal,letter,oneside,twocolumn]{IEEEtran}

\usepackage[noadjust]{cite}
\usepackage{srcltx}
\usepackage{psfrag}
\usepackage{epsfig}
\usepackage[tbtags,sumlimits,nointlimits,reqno]{amsmath}
\usepackage{amssymb}
\usepackage{xcolor}
\usepackage{float}
\usepackage{subfigure} 
\usepackage{soul}
\usepackage{footnote}

\setstcolor{red}

\usepackage{showkeys}   

\begin{document}

\title{Magneto-Electric Dipole Antenna Arrays}

\author{%
       Shulabh~Gupta,~\IEEEmembership{Member,~IEEE,} Li Jun Jiang,~\IEEEmembership{Senior 	Member,~IEEE,} and~Christophe~Caloz,~\IEEEmembership{Fellow,~IEEE}
\thanks{S. Gupta and L. J. Jiang are with the Electrical and Electronic Engineering Department of University of Hong Kong, China (SAR) Email: shulabh@hku.hk.}
\thanks{C. Caloz is with the Department of Electrical
Engineering, PolyGrames Research Center, \'{E}cole Polytechnique de Montr\'{e}al,
Montr\'{e}al, QC, Canada.}
}
\markboth{IEEE Transactions on Antenna and Propagation, 2014}{Shell \MakeLowercase{\textit{et al.}}: Bare Demo of IEEEtran.cls for
Journals}

\maketitle

\begin{abstract}
A planar magneto-electric (ME) dipole antenna array is proposed and demonstrated by both full-wave analysis and experiments. The proposed structure leverages the infinite-wavelength propagation characteristic of composite right/left-handed (CRLH) transmission lines to form high-gain magnetic radiators combined with radial conventional electric radiators, where the overall structure is excited by a single differential feed. The traveling-wave type nature of the proposed ME-dipole antenna enables the formation of directive arrays with high-gain characteristics and scanning capability. Peak gains of 10.84~dB and 5.73~dB are demonstrated for the electric dipole and magnetic-dipole radiation components, respectively.
\end{abstract}


%
\IEEEpeerreviewmaketitle
%
\section{Introduction}

The concept of collocated magnetic and electric dipole radiators (M- and E-dipoles) has recently attracted a considerable attention due to the increased demand for multi-functionality and switchable characteristics in modern communication systems \cite{ME_Review}. Applications include anti-collisions systems for vehicular transport~\cite{LoopArray_Vehicular}, MIMO systems for high-speed communication \cite{EM_MIMO}, enhanced polarization diversity MIMO systems~\cite{Sam_PD_CRLHDIFF}, for instance. While conventional phased array antennas have been sporadically used in such applications, \cite{Loop_Array, Loop_Array_old, PhasedArray}, a clever combination of the two fundamental types of radiators, namely magnetic and electric dipoles, into a Magneto-Electric (ME) antenna, may offer enhanced performance functionalities.  

The concept of ME antennas was introduced in~\cite{Equal_E_H} where identical E- and H-plane radiation patterns are obtained by exciting simultaneously an electric dipole and a magnetic dipole. This concept was further explored in slot configurations in \cite{Slot_Dipole} and analyzed for an array in \cite{Colocated} and recently for UWB applications in~\cite{Luk_EM, Luk_ME} for resonant-type configurations. These ME antennas are non-planar and are difficult to fabricate. A planar ME monopole configuration based on composite right/left-handed (CRLH) transmission lines was proposed in~\cite{CRLH_Mono, Caloz-MTM-Book}, but it requires a dual-feeding mechanism to excite the E- and M-radiators separately. Moreover being a resonant-type (as opposed to a traveling-wave type), it cannot be used to form a single-feed array.

In this work, an ME-dipole antenna structure is proposed in a planar configuration based on CRLH transmission lines, where the electric and magnetic radiators are excited simultaneously using a single differential feed. While E- and M-radiators are conventionally used in a superposed configuration where the two radiators enhance each other to form a single radiating element \cite{Luk_EM, Luk_ME}, the proposed structure acts as two distinct antennas in one structure with their individual radiation characteristics. Furthermore, compared to conventional resonant-type ME antennas, the proposed antenna is travelling-wave in nature, which makes it suitable for forming simple high-gain ME-dipole antenna arrays with dual-polarized radiation characteristics \cite{DualPolCRLH1,DualPolCRLH2}. Finally, due to the CRLH leaky-wave property of the structures, the proposed structure offers frequency beam-scanning of both E- and M-dipoles.

The paper is organized as follows. Section II introduces the concept of ME-dipole antennas based on CRLH structures. Using a typical implementation of a CRLH based structure, the radiation characteristics of the proposed ME-dipole antenna is described in details and modelled using an array-factor theory to provide further insight into the structure. Section III  presents the experimental prototypes of the ME-dipole antenna array and its beam-switching characteristics. Finally, the conclusions are provided in Section IV.

\section{Magneto-Electric (ME) Dipole Antenna Array}

\subsection{Structure and Principle}

\begin{figure}[htbp]
\begin{center}
\psfrag{+}[c][c][0.75]{$+$}
\psfrag{-}[c][c][0.75]{$-$}
\psfrag{z}[c][c][0.75]{$Z_0$}
\psfrag{a}[c][c][0.75]{$L_R/2$}
\psfrag{b}[c][c][0.75]{$2C_L$}
\psfrag{c}[c][c][0.75]{$C_R$}
\psfrag{d}[c][c][0.75]{$L_L$}
\includegraphics[width=\columnwidth]{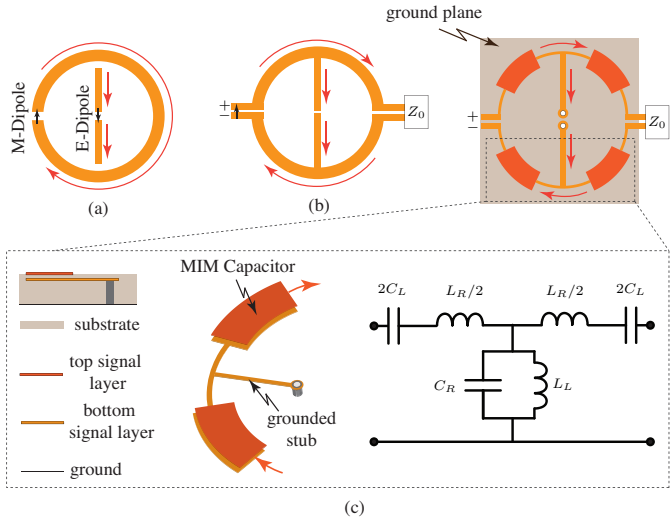}
\caption{Proposed planar magneto-electric (ME) antenna. a) Co-located E- and M-dipoles with separate feeds. b) 2-port travelling-wave configuration consisting of merged co-located E- and M-dipoles with a common differential feed. c) Composite right/left-handed (CRLH) implementation of the ME-dipole operating in the infinite wavelength regime and radiating into the half-space above the ground plane.} \label{Fig:Concept}
\end{center}
\end{figure}

Consider the structure of Figure~\ref{Fig:Concept}(a) consisting of a magnetic loop dipole and an electric-wire dipole placed within it along the radial direction, where each of the two dipoles has its own and independent differential excitation. When the two dipoles are excited simultaneously, combined magnetic  and dipole radiation is achieved in the far-field of the structure. Such a combined radiator is referred to as an ME dipole antenna. Although being straight-forward, this configuration is unpractical because it requires two separate and intertwined feeds. However, this antenna can be replaced by the merged structure shown in Fig.~\ref{Fig:Concept}(b) with $Z_0=0$ (direct loop connection), where the ring is differentially excited from two antipodal points, where part of the loop current is used to drive the electric dipole at the centre of the ring. This structure is more practical since it uses a \textit{single differential excitation}. Moreover, if a second differential port, of impedance $Z_0\ne0$, is placed at the other side of the loop, the structure transforms from a \textit{resonant-type} to a \textit{traveling-wave} type antenna.

It is well-known that the gain of a magnetic loop dipole is small but can be theoretically improved by increasing the size of the loop while maintaining the current along the loop constant \cite{Balanis-Antenna-Book}. This can be practically achieved by using a composite right/left-handed (CRLH) transmission line structure operated in its infinite-wavelength regime \cite{Caloz-MTM-Book, MTM_Eleft_Book, Capolino_Eleft_Book}. 

A CRLH transmission line is a periodic structure whose unit-cell consists of a series capacitance, $C_L$, and a shunt inductance, $L_L$, in addition to a series inductance, $L_R$, and shunt capacitance, $C_R$, as shown in the circuit model of Fig.~\ref{Fig:Concept}(c) \cite{Caloz_MT_2009}. A common implementation of a CRLH transmission line is the metal-insulator-metal (MIM) implementation also illustrated in the figure. The parallel-plate MIM capacitor provides $C_L$ and the shunt stub connected to ground using a conducting via, provides $L_L$. The other elements, $L_R$ and $C_R$, are modelled using a series transmission line section and a shunt stub, respectively. Such an artificial transmission line acts as a left-handed (LH) transmission line at low frequencies and a right-handed (RH) transmission line at high frequencies. It has been extensively applied in realizing both guided-wave and radiative components, including phase-shifters \cite{MTM2,MTM3}, tight-couplers, leaky-wave and phased-array antennas \cite{Caloz-MTM-Book,PhasedArray}, among many other applications. A CRLH transmission line is characterized by the following dispersion relation \cite{Caloz-MTM-Book}

\begin{subequations}
\begin{equation}
\beta(\omega) = \frac{1}{p}\cos^{-1}\left(1 - \frac{\kappa}{2}\right)
\end{equation}

where

\begin{equation}
\kappa=\left(\frac{\omega}{\omega_R}\right)^2 + \left(\frac{\omega_L}{\omega}\right)^2 - \kappa\omega_L^2,
\end{equation}
\end{subequations}

\noindent with $\kappa= L_LC_r + L_RC_L$, $\omega_R=1/\sqrt{L_RC_R}$ and $\omega_L=1/\sqrt{L_LC_L}$, which corresponds to the dispersion relation of the CRLH radiation space harmonic, the harmonic $n=0$, featuring an infinite wavelength at the transition frequency between the LH and RH bands \cite{Caloz-MTM-Book, Caloz_MT_2009}. Depending on the relative values of the LH and RH contributions, this transmission line can be unbalanced or balanced, i.e. exhibiting or not a gap between the two bands. The balanced condition requires that the individual elements are designed such that $L_LC_R = L_RC_L$, which is also the condition for broad band matching \cite{Caloz-MTM-Book}. Under the balanced condition, the CRLH propagation constant simplifies to

\begin{equation}
\beta(\omega) = \left(\frac{\omega}{\omega_R} - \frac{\omega_L}{\omega}\right).
\end{equation}

\noindent The frequency, where the dispersion curve passes from the LH band to the RH band, $\beta = 2\pi/\lambda_g = 0$, is referred to as the transition frequency. At this frequency, the structure supports an infinite-wavelength ($\lambda_g=\infty$) traveling-wave ($v_g = d\omega/d\beta \ne 0$) regime. This $\beta=0$ regime has some similarity with the cutoff regime in a waveguide, but with the essential difference that it supports a traveling wave exactly at $\beta=0$. Consequently, the phase of the fields on the structure is macroscopically uniform along the CRLH transmission line structure, irrespectively of its length. Moreover, a CRLH transmission line can also be operated in a radiative leaky-wave mode when open to free space, since the CRLH dispersion curve penetrates into the fast-wave region, where the phase velocity $v_p(\omega) = \omega/\beta(\omega)$ is larger than the speed of light \cite{Caloz-MTM-Book, CRLH_Loop, Zeroth_order_CRLH_Loop}. The resulting leaky-wave antenna (LWA) radiates from backfire to endfire including broadside as frequency is scanned from backward to forward regions \cite{Leaky_Book, Leaky_jackson}. It is to be noted that rigorous conditions exist to achieve a seamless transition from the LH band to the right-hand band, to ensure that the resulting CRLH structure is balanced, without any gain-drop at broadside while scanning from backward to forward region \cite{Otto_Leaky}.

\begin{figure*}[htbp]
\begin{center}
\subfigure[]{
\psfrag{a}[c][c][0.75]{$Z_0$}
\psfrag{x}[c][c][1]{$x$}
\psfrag{y}[c][c][1]{$y$}
\psfrag{z}[c][c][1]{$z$}
\psfrag{+}[c][c][0.75]{$+$}
\psfrag{-}[c][c][0.75]{$-$}
\includegraphics[width=0.6\columnwidth]{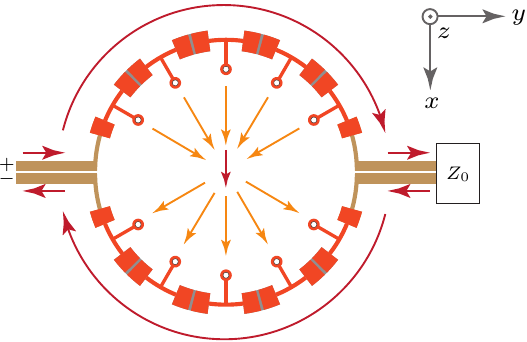}}
\subfigure[]{
\psfrag{a}[c][c][0.8]{phase constant $\phi[S_{21}]$ (rad/m)}
\psfrag{b}[c][c][0.8]{frequency (GHz)}
\psfrag{c}[c][c][0.8]{S-parameters (dB)}
\psfrag{d}[c][c][0.8]{$f_0$}
\psfrag{e}[c][c][0.8]{$-k$}
\psfrag{f}[c][c][0.8]{$+k$}
\psfrag{x}[c][c][0.6]{RH $f>f_0$}
\psfrag{y}[c][c][0.6]{LH $f<f_0$}
\includegraphics[width=0.9\columnwidth]{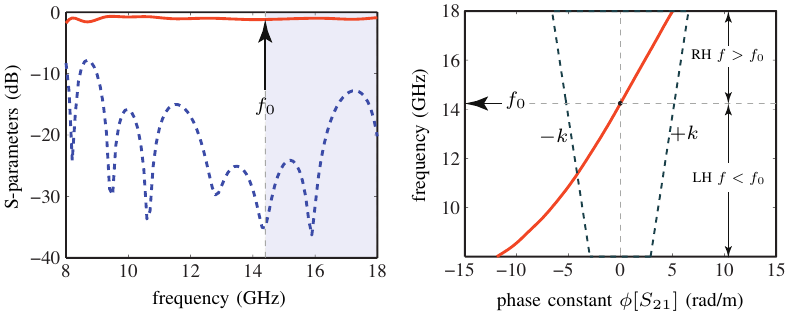}}
\subfigure[]{
\psfrag{a}[l][c][0.6]{$G_\phi$ at $\phi=0^\circ$}
\psfrag{b}[l][c][0.6]{$G_\theta$ at $\phi=0^\circ$}
\psfrag{c}[l][c][0.6]{$G_\phi$ at $\phi=90^\circ$}
\includegraphics[width=0.4\columnwidth]{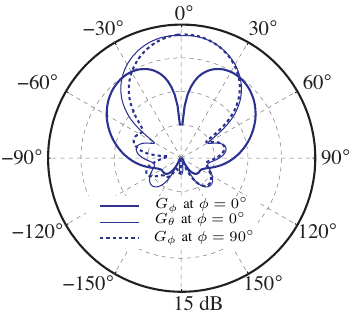}}
\caption{Proposed ME dipole antenna based on CRLH transmission lines operated in the infinite wavelength regime, where the size of the ring, in principle, can made arbitrarily large to enhance the gain. a) Structure layout. b) Typical S-parameters and its corresponding dispersion diagram, and d) typical radiation patterns at broadside, computed using FEM-HFSS. The physical parameters used are the same as in Fig.~\ref{Fig:ArrayHFSS}. The $G_\theta$ component at $\phi=90^\circ$ is negligible.} \label{Fig:Concept_HFSS}
\end{center}
\end{figure*}

These unique properties can be leveraged to transform the ME-dipole structure of Fig.~\ref{Fig:Concept}(b) to that shown in Fig.~\ref{Fig:Concept_HFSS}(a), where the series elements ($L_R$ and $C_L$) of the CRLH transmission line form the current loop, i.e. the M-dipole, and the shunt elements ($L_L$ and $C_R$) form the E-dipole. Then, owing to the infinite-wavelength behaviour of the CRLH transmission line at the transition frequency, the size of the ring can be made arbitrarily large without altering the current phase or wave-propagation direction. Consequently, compared to the conventional resonant loop antennas, the operation frequency of the CRLH loop does not depend on the size of the ring, but on the transition frequency depending on the unit-cell design of the CRLH structure. The M-dipole gain can be progressively increased by increasing the size of the current loop, as shown in Fig.~\ref{Fig:Concept_HFSS}(a), with a larger number of unit cells while maintaining infinite-wavelength propagation at the design frequency. The circular array of shunt stubs contributes to the E-dipole contribution where all the radial dielectric dipoles have cancelling parts in the horizontal direction, to generated a global vertical dipole, as shown in Fig.~\ref{Fig:Concept}(d). The E-dipole gain thus depends on the shunt stubs but cannot be independently controlled due to the CRLH  balancing requirement. It is to be noted that compared to the ME-dipole of Fig.~\ref{Fig:Concept}(b), the CRLH based implementation typically requires a ground plane restricting radiation to half-space in contrast to the bi-directional radiation of conventional dipole antennas.

Figure~\ref{Fig:Concept_HFSS}(b) shows typical S-parameters and the dispersion diagram of such a CRLH structure. Good matching is clearly apparent across a wide frequency band, including the transition frequency $f_0$. A seamless transition from the LH to the RH band is clearly observed in the dispersion diagram in Fig.~\ref{Fig:Concept_HFSS}(b). The part of the dispersion curve crossing the triangular fast-wave region, corresponds to leaky-wave radiation. In particular, the transition frequency $f_0$ points to the broadside direction \cite{Caloz_MT_2009}.

\subsection{Features and Benefits}

The proposed antenna offers several benefits compared to the conventional ME-dipole antennas:

\begin{enumerate}

\item It is easy to fabricate, due to its \emph{planar configuration}, and is compatible with MMIC technology.
\item It has a simple configuration requiring only a \emph{single differential feed} to excite both the E- and M-dipoles.
\item As opposed to resonant-type ME-dipole antennas, the proposed structure is of \emph{traveling-wave type}, and therefore can be extended to a single feed array configuration for high-gain and scanning performance.
\item The proposed antenna is \emph{multi-functional}, capable of frequency-scanning in two orthogonal planes with orthogonally polarized fields components. Fixed-frequency beam-scanning can naturally be achieved using voltage controlled varactors integrated in the antenna \cite{Caloz-MTM-Book}.
\item Due to its intrinsic differential nature, the proposed antenna is compatible with \emph{high-density circuits} \cite{Diff_yang} and, in particular, can be naturally integrated with differential amplifiers for \emph{beam-shaping} and \emph{beam-forming} applications \cite{Diff_Amps}\cite{CRLH_Active}\cite{CRLH_Distributed_Amplifier}.
	
\end{enumerate}

\begin{figure*}[htbp]
\begin{center}
\subfigure[]{
\psfrag{z}[c][c][0.8]{$Z_0$}
\psfrag{x}[c][c][1]{$x$}
\psfrag{m}[c][c][1]{$z$}
\psfrag{y}[c][c][1]{$y$}
\psfrag{a}[c][c][0.8]{$a$}
\psfrag{b}[c][c][0.8]{$p$}
\psfrag{c}[c][c][0.8]{$\Delta \theta_1$}
\psfrag{d}[c][c][0.8]{$\Delta \theta_2$}
\psfrag{e}[c][c][0.8]{$\Delta \theta_3$}
\psfrag{f}[c][c][0.8]{$\ell_s$}
\psfrag{h}[c][c][0.8]{$w_s$}
\psfrag{g}[c][c][0.8]{$w_\text{st}$}
\includegraphics[width=2\columnwidth]{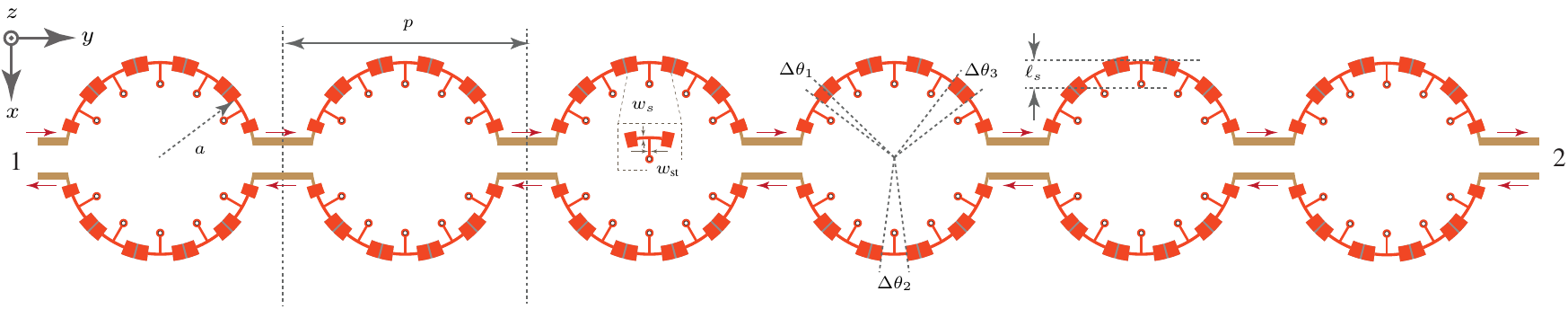}}
\subfigure[]{
\psfrag{a}[c][c][1]{$G_y(\theta,\phi)$}
\psfrag{b}[c][c][1]{$G_x(\theta,\phi)$}
\psfrag{e}[l][c][0.8]{$f_1 = 14$~GHz}
\psfrag{f}[l][c][0.8]{$f_2 = 14.25$~GHz}
\psfrag{h}[l][c][0.8]{$f_3 = 15$~GHz}
\psfrag{x}[c][c][1]{$x$}
\psfrag{y}[c][c][1]{$y$}
\includegraphics[width=1.8\columnwidth]{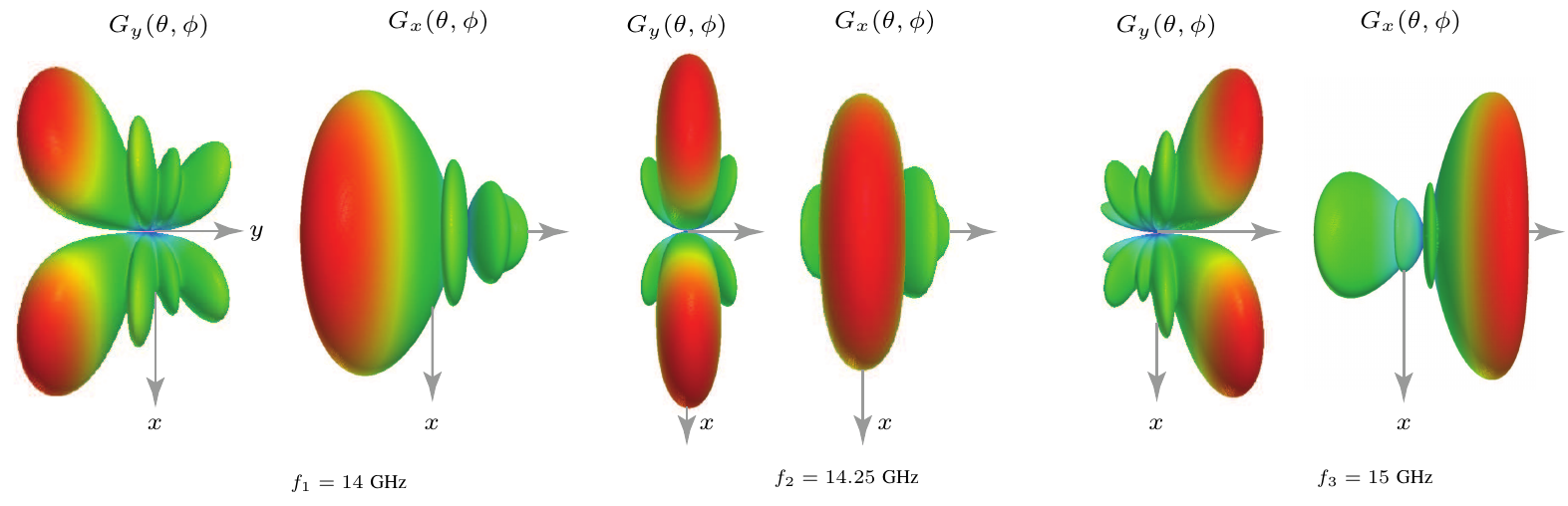}}
\subfigure[]{
\psfrag{b}[c][c][1]{\shortstack{$G_\phi(\theta)$, $\phi=0^\circ$ \\M-dipole}}
\psfrag{a}[c][c][1]{\shortstack{$G_\theta(\theta)$, $\phi=0^\circ$\\E-dipole}}
\psfrag{c}[c][c][1]{\shortstack{$G_\phi(\phi)$, $\theta=90^\circ$\\M-dipole Array}}
\psfrag{d}[c][c][1]{\shortstack{$G_\phi(\theta)$, $\phi=90^\circ$\\E- and M-dipole Array}}
\psfrag{e}[l][c][0.8]{$f_\text{LF} = 14.0$~GHz}
\psfrag{f}[l][c][0.8]{$f_0 = 14.25$~GHz}
\psfrag{h}[l][c][0.8]{$f_\text{RF} = 14.5$~GHz}
\includegraphics[width=2\columnwidth]{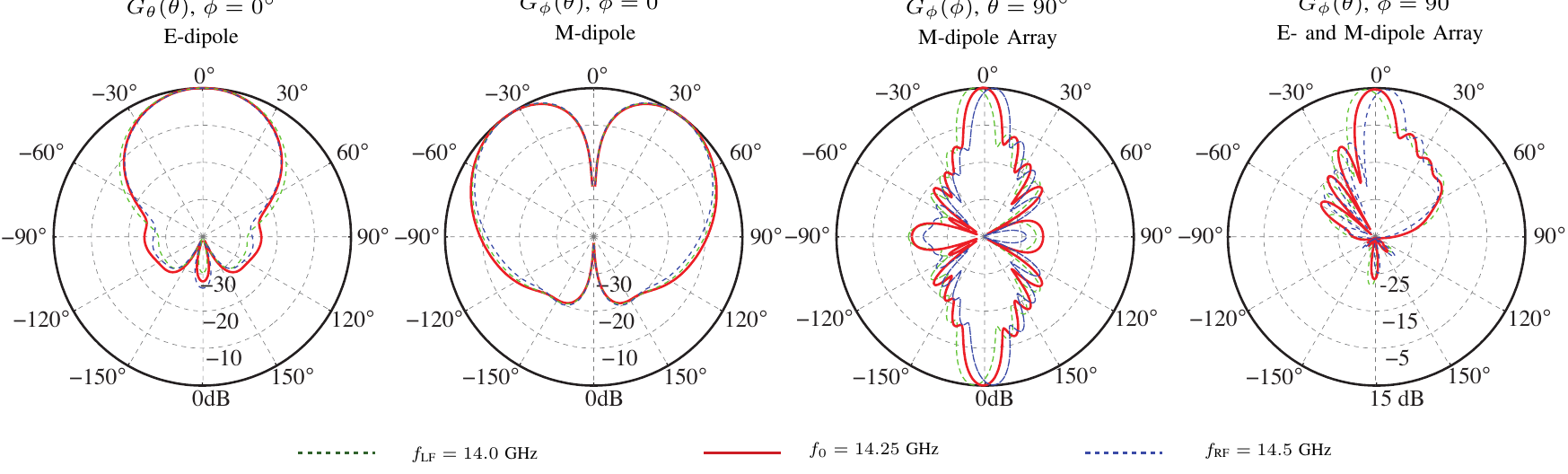}}
\subfigure[]{
\psfrag{a}[c][c][0.8]{M-Dipole $G_\phi(\theta, \phi=0^\circ)$ dB}
\psfrag{b}[c][c][0.8]{E-Dipole $G_\theta(\theta, \phi=0^\circ)$ dB}
\psfrag{c}[c][c][1]{Elevation angle $\theta$ ($^\circ$)}
\psfrag{d}[c][c][1]{frequency (GHz)}
\psfrag{e}[c][c][0.8]{Gain (dB)}
\psfrag{f}[c][c][0.6]{$f_0$}
\psfrag{m}[c][c][1]{S-parameters (dB)}
\psfrag{n}[c][c][1]{$\phi[S_{21}]$ (deg)}
\psfrag{p}[c][c][0.8]{$|S_{11}|$}
\psfrag{q}[c][c][0.8]{$S_{21}$}
\psfrag{g}[c][c][0.8]{$G_\phi(\theta = \pm32^\circ, \phi=0^\circ)$}
\psfrag{h}[c][c][0.8]{$G_\theta(\theta = 0^\circ, \phi=0^\circ)$}
\includegraphics[width=2\columnwidth]{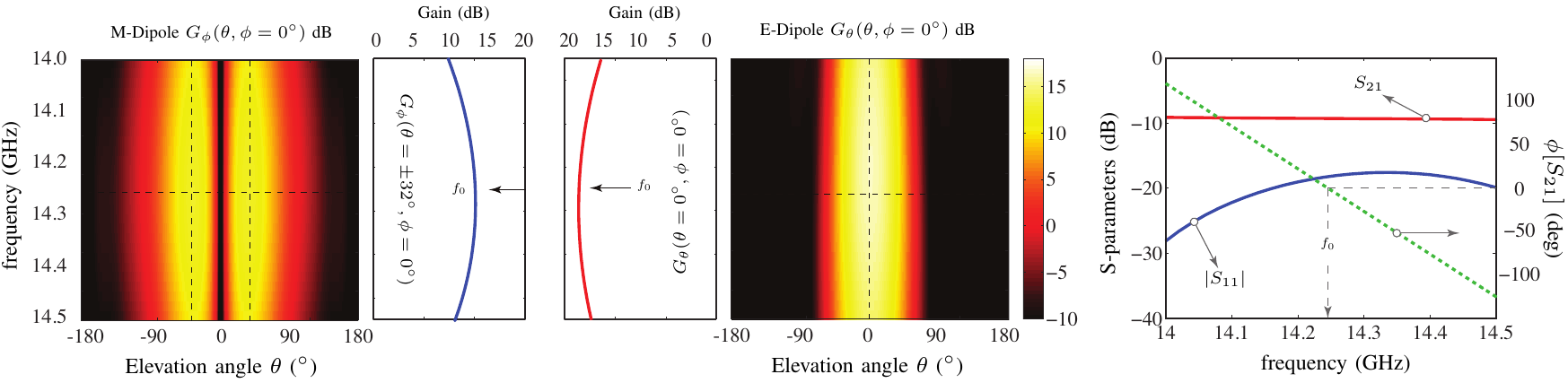}}
\caption{Proposed ME-dipole antenna array. a) Structure layout with each CRLH unit cell implemented in metal-insulator-metal (MIM) technology. b) Typical 3-D normalized radiation patterns at the transition frequency $f_0=14.25$~GHz of the CRLH structure and the frequencies, $f_\text{LF} = 14.0$~GHz and $f_\text{RF}=15.0$~GHz in the left- and right-handed bands, respectively, computed using FEM-HFSS. c) Normalized radiation patterns in three co-ordinate planes. d) Typical gain-pattern maps showing the M- and E-dipole gains along with the 2-port S-parameters. The structure consists of 2 substrates (Taconic RF-30 $\varepsilon_r = 3$) of height, $0.508$~mm and $0.127$~mm, respectively. The structural parameters are: $p=16$~mm, $a=7$~mm, $w_s = 0.254$~mm, $w_\text{st}=0.254$~mm, $\ell_s = 2$~mm, $\Delta\theta_1=6^\circ$, $\Delta\theta_2 =\Delta\theta_3=15^\circ$. The via radius is $0.254$~mm and the ground plane width is $6a$.} \label{Fig:ArrayHFSS}
\end{center}
\end{figure*}

\subsection{Antenna Array Configuration}

As mentioned in the previous sections, placing a second differential port on the current loop of Fig.~\ref{Fig:Concept}(b-d) converts the ME-dipole antenna into a \textit{traveling-wave type antenna}. This enables the realization of an ME-dipole antenna array, as shown in Fig.~\ref{Fig:ArrayHFSS}(a), consisting of $N=6$ rings\footnote{The choice of $N=6$ rings is to ensure that the structure can be fabricated using the multi-layer process at our facility which is restricted to 12.7~cm$ \times$ 6.4~cm of area.}, offering the possibilities for enhanced gain and directivity performance. This structure may be seen as a phased array of electric and magnetics dipole antennas, where the radiation mechanism is leaky-wave in nature and follows CRLH dispersion \cite{Caloz_AWPL}.

The typical radiation patterns of such an array is shown in Fig.~\ref{Fig:ArrayHFSS}(b-c) for the transition frequency $f_0=14.25$~GHz and frequencies $f_{LH}$ and $f_{RH}$ in the LH-band and RH-band of the CRLH transmission line, respectively. At this frequency the ME-dipole array is about $4.6\lambda_0$ in the $y-$direction, and the corresponding CRLH structure is $6.5\lambda_0$ following the circumference of the rings. Fig.~\ref{Fig:ArrayHFSS}(b) shows the the components $E_x$ and $E_y$ for each frequency and Fig.~\ref{Fig:ArrayHFSS}(c) shows the corresponding 2D patterns in three principal cuts.  The following observations can be made from these patterns:

\begin{enumerate}
\item The $E_y$ field component corresponds to the M-dipole contribution only and exhibits a typical radiation pattern of a loop antenna over a ground plane with a null at $\theta = 0^\circ$.
\item The $E_x$ field component corresponds primarily to the E-dipole contribution and exhibits a minimum at $\theta = \pm90^\circ$. 
\item A full-space scanning from backward to forward regions including broadside is achieved for both E- and M-dipoles, as typical in CRLH leaky-wave antennas.
\item While the E-dipole components $E_\phi$ scans in the $\phi = 90^\circ$ plane, the M-dipole scans in the $\theta = 90^\circ$ plane with the $E_\phi$ polarization.
\item While the E-dipole pattern has a single beam in $+z$-space, the M-dipole pattern exhibits two beams, one in each positive $x-z$ plane quadrants. 
\item There is only one dominant field component, $E_\phi$, in both $x-y$ and $y-z$ planes, along with zero cross-polarization $E_\theta$.
\end{enumerate}

\begin{figure}[htbp] 
 \begin{minipage}[c]{0.5\columnwidth}
\psfrag{a}[c][c][0.8]{Frequency (GHz)}
\psfrag{b}[c][c][0.7]{accepted power, $P_\text{acc}$}
\psfrag{d}[c][c][0.7]{radiated power, $P_\text{rad}$}
\psfrag{c}[c][c][0.7]{efficiency $\eta$ }    
\includegraphics[width=\textwidth]{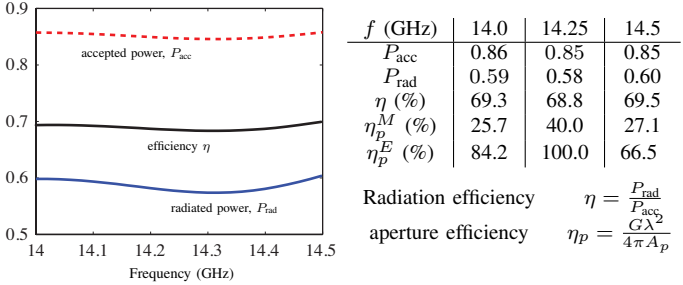}
\end{minipage}
\begin{minipage}[c]{0.4\columnwidth}
\footnotesize{\begin{tabular}{c|c|c|c}
$f$ (GHz) & 14.0 & 14.25 & 14.5\\
\hline
$P_\text{acc}$ & 0.86 & $0.85$ &0.85 \\
$P_\text{rad}$ & $0.59$ & 0.58 &0.60 \\
$\eta$ (\%) & 69.3 & 68.8 & 69.5 \\
$\eta_p^M$ (\%)& 25.7 & 40.0 & 27.1\\
$\eta_p^E$ (\%)& 84.2 & 100.0 & 66.5
\vspace{2mm}
\end{tabular}}
\centering
\footnotesize{\begin{tabular}{cc}
Radiation efficiency & $\eta = \frac{P_\text{rad}}{P_\text{acc}}$ \\
aperture efficiency & $\eta_p = \frac{G\lambda^2}{4\pi A_p}$
\vspace{2mm}
\end{tabular}}
\end{minipage}
\caption{Various radiation performance parameters of the ME-dipole array of Fig.~\ref{Fig:ArrayHFSS}, computed using FEM-HFSS.}\label{Fig:EfficiencyHFSS}
\end{figure}

\noindent The S-parameters and gain performance of this array design is shown in Fig.~\ref{Fig:ArrayHFSS}(d). Good match is obtained around the transition frequency $f_0$, with a seamless transition from the LH to RH frequency band, as clearly evident from the corresponding continuous transmission phase across $f_0$, also shown in Fig.~\ref{Fig:ArrayHFSS}(d). Both E- and M-dipoles exhibit a high peak gain greater than 10~dB within the bandwidth of interest around the transition frequency. For example, a substantial increase in the gain at broadside is achieved compared to that of a single-ring structure, as seen from Fig.~\ref{Fig:ArrayHFSS}(c). The gain map of the M-dipole shows that the maximum gain at individual frequencies occur at about $\theta = \pm32^\circ$ due to the presence of the finite-size ground plane, instead of $\theta=\pm90^\circ$. Across the frequency range, the transition frequency exhibits the highest gain, for both E- and M-dipole radiation components, with a smooth drop in gain on either side of it. This is expected as the uniform current distribution is only achieved at $f_0$. At other frequencies, the current distribution deviates from the uniform distribution, therefore the gain subsequently drops due to spatial dispersion. This sensitivity of the current distribution across the ring structure makes the proposed ME-dipole antenna \textit{narrowband}. Fig.~\ref{Fig:EfficiencyHFSS} summarizes the performance of the antenna of Fig.~\ref{Fig:ArrayHFSS}. A fairly constant antenna efficiency of about $69\%$ is seen across the bandwidth of interest.

\begin{figure}[htbp]
\begin{center}
\psfrag{x}[c][c][1]{$x$}
\psfrag{z}[c][c][1]{$z$}
\psfrag{y}[c][c][1]{$y$}
\psfrag{a}[l][c][0.8]{$z_0 = 10$~mm}
\psfrag{b}[l][c][0.8]{$z_0 = 5$~mm}
\psfrag{c}[c][c][1]{distance $y$ (mm)}
\psfrag{d}[c][c][1]{E-Field magnitude $|E(y)|$ (dB)}
\psfrag{e}[c][c][1]{20~mm/division}
\psfrag{f}[c][c][1]{10~mm/division}
\psfrag{m}[c][c][0.8]{\color{white}($x=0$)\color{black}}
\psfrag{n}[c][c][0.8]{\color{white}($z_0=10$)~mm\color{black}}
\includegraphics[width=0.8\columnwidth]{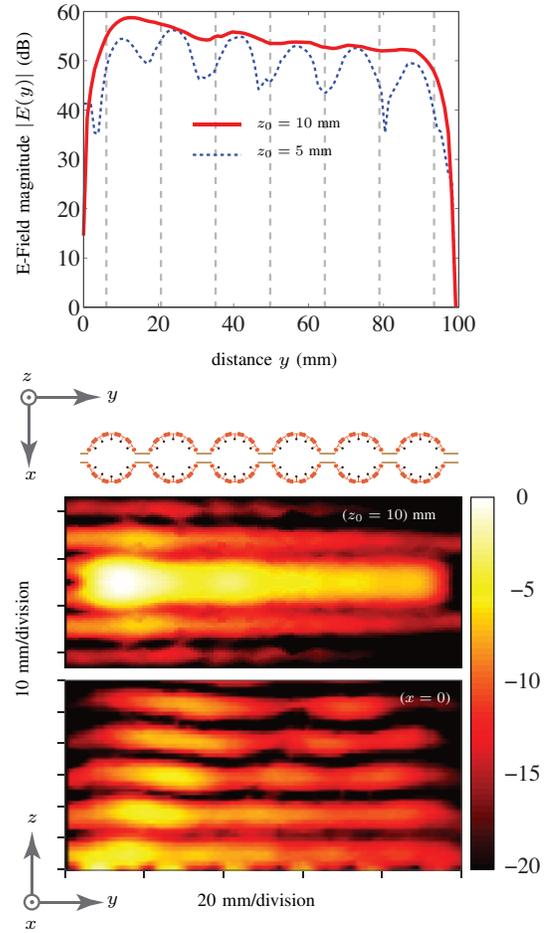}
\caption{Full-wave (FEM-HFSS) computed near fields just above the ME-dipole array of Fig.~\ref{Fig:ArrayHFSS}(a) at the transition frequency $f_0=14.25$~GHz, demonstrating the $\beta=0$ operation of the CRLH structure.}\label{Fig:NFHFSS}
\end{center}
\end{figure}

Figure~\ref{Fig:NFHFSS} shows full-wave computed E-field plots in the near-field region of the antenna, in the two orthogonal planes $z=10$~mm and $x=0$, at the transition frequency $f_0$. As expected, a uniform field profile is observed across the structure, confirming the $\beta=0$ operation of the CRLH structure. The decrease in the field amplitude along the $y-$axis is due to the leaky-wave radiation loss and the power dissipation in the conductor and the dielectric. The corresponding E-field distribution along the antenna, $|E(y)|$, at two different locations above the antenna, shows that the plane-waves start to form from about $z=10$~mm, which corresponding to roughly half-wavelength distance from the antenna, i.e. $\lambda_0/2$ at $f_0$.

 \subsection{Comparison with an Array of Ideal Dipole Radiators}

\begin{figure}[htbp]
\begin{center}%
\subfigure[]{
\psfrag{a}[c][c][0.7]{$p$}
\psfrag{b}[c][c][0.7]{$2r_0$}
\psfrag{c}[c][c][0.7]{$\ell$}
\psfrag{x}[c][c][1]{$x$}
\psfrag{y}[c][c][1]{$y$}
\psfrag{z}[c][c][1]{$z$}
\includegraphics[width=\columnwidth]{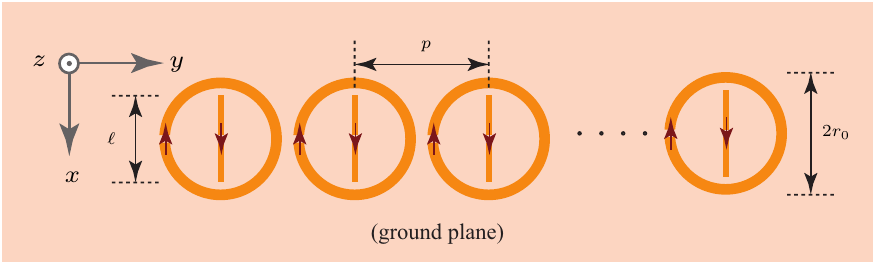}}
\subfigure[]{
\psfrag{b}[c][c][0.7]{\shortstack{$G_\phi(\theta)$, $\phi=0^\circ$ \\($x-z$ plane)}}
\psfrag{a}[c][c][0.7]{\shortstack{$G_{\phi, \theta}(\theta)$, $\phi=0^\circ$ \\($x-z$ plane)}}
\psfrag{d}[c][c][0.7]{\shortstack{$G_{\phi, \theta}(\theta)$, $\phi=90^\circ$ \\ ($y-z$ plane)}}
\psfrag{x}[c][c][0.8]{\shortstack{Proposed ME-Dipole \\Array Fig.~\ref{Fig:ArrayHFSS}(a)}}
\psfrag{y}[c][c][0.8]{\shortstack{Ideal array, \\$2\pi r_0 \approx  \lambda_0/3$}}
\psfrag{z}[c][c][0.8]{\shortstack{Ideal array, \\$2\pi r_0 = 2\pi a = 2.2\lambda_0$}}
\psfrag{c}[l][c][0.8]{$G_\phi$}
\psfrag{e}[l][c][0.8]{$G_\theta$}
\psfrag{w}[l][c][0.8]{$a$}
\psfrag{v}[l][c][0.8]{$p$}
\includegraphics[width=\columnwidth]{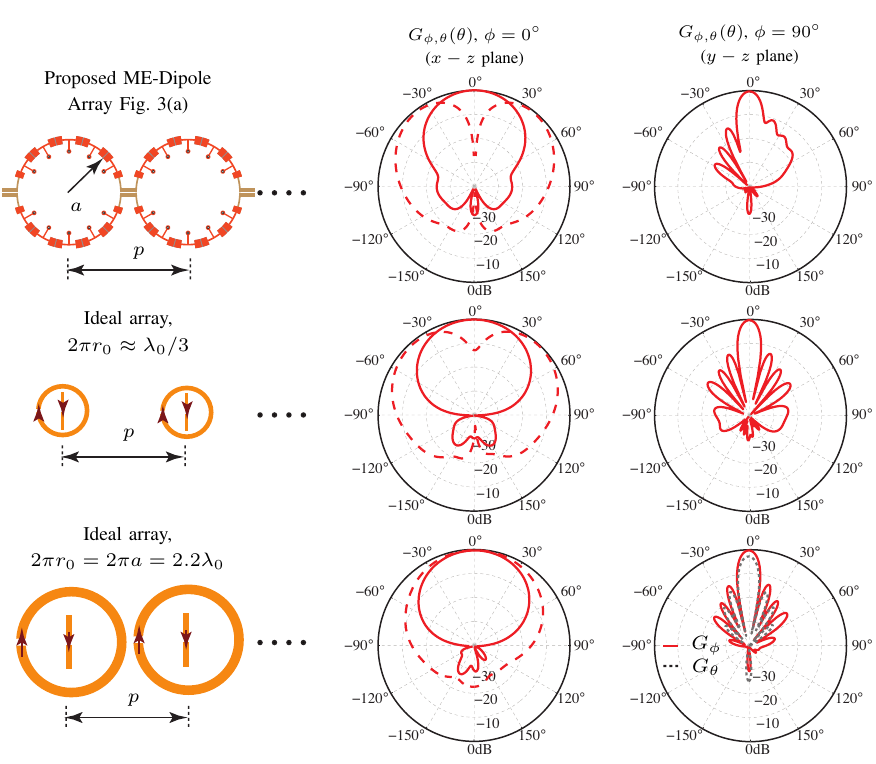}}
\caption{Radiation patterns comparisons between the proposed and ideal ME-Dipole array. a) An ideal array consisting of co-located current loops and electric-dipoles, with their own signal inputs. b) Full-wave (FEM-HFSS) radiation patterns of the proposed ME-dipole array of Fig.~\ref{Fig:ArrayHFSS} compared with an ideal array formed using small current loops $r_0<< \lambda_0$, and current loop of identical size $r_0=a$.} \label{Fig:ArrayHFSSvsIdeal}
\end{center}
\end{figure}

As mentioned above, the proposed CRLH ME-dipole antenna, based on CRLH structure, may be seen as an array of uniform current loops
 loop of uniform current, collocated with short electric dipoles, as shown in Fig.~\ref{Fig:Concept_HFSS}. To confirm the validity of this equivalence, an array of ideal current loops and dipoles is considered, as shown in Fig.~\ref{Fig:ArrayHFSSvsIdeal}(a). This array consists of $N=6$ differentially excited rings of radius $r_0$ and circumference $C$, along with collocated short electric-dipoles of length $\ell$, with their own differential excitations, on top of a grounded substrate. The separation $p$ between the radiating units is taken the same as used in Fig.~\ref{Fig:ArrayHFSS} for fair comparison. Furthermore, all the radiating elements are simultaneously excited to emulate the broadside operation of the array. Consequently, all the forthcoming comparisons are made at the transition frequency, with $\beta=0$, of the ME-dipole array, i.e. $f_0=14.25$~GHz.

{Figure~\ref{Fig:ArrayHFSSvsIdeal}(b) shows the comparison of the simulated radiation patterns of the ideal structure with the proposed ME-dipole array, under two different conditions: 
\begin{enumerate}
\item Case I $C \approx 0.67\lambda_0$: In this case, the conventional loop antennas have a close to uniform current distribution \cite{Balanis-Antenna-Book}. The following observations can be made:

\begin{itemize}
\item A broadside null is achieved in the magnetic-loop component (i.e. $G_\phi$ in $\phi=0^\circ$) of both ideal and CRLH ME-dipole arrays.
 While the ideal loops are electrically small, the CRLH array has a much larger loop circumference ($C\approx 2.2\lambda_0$). This further confirms that the ME-dipole array loop has a uniform current distribution, although having an electrically large size.
 \item There is no $G_\theta$ component in the $y-z$ plane in both cases, since the electric dipole is orthogonally polarized in both cases. 
 \item A typical electric-dipole response is seen in $x-y$ plane in both cases, with a maximum at broadside. In addition, a directive beam is formed in the $y-z$ plane due to the presence of a co-polarized electric-dipole array. This confirms that the circular array of radiating stubs in each loop of the ME-dipole array indeed acts as a single electric-dipole located at the centre of the ring.
\end{itemize}
\item Case II $C\approx 2.1$: In this case, the loop is electrically large with a a non-uniform field distribution. Two observations can be made:
\begin{itemize}
\item A broadside maximum is observed in the magnetic-loop component (i.e. $G_\phi$ in $\phi=0^\circ$) of the ideal array, as opposed to the null in the case of CRLH ME-dipole array.
\item There is strong cross-polarization component $G_\theta$ in the $y-z$ plane which does not exist in the proposed CRLH array.
\end{itemize}

They demonstrate that, a very different radiation pattern is obtained compared to that of the proposed structure or the ideal structure consisting of electrically small dipoles.
\end{enumerate}
}

It is to be noted that the ideal array structure is only used here for qualitative comparisons of the radiation patterns, to validate its link with the CRLH ME-dipole array. These ideals arrays, otherwise, are inefficient radiators, both due to their sizes, and a presence of a ground plane, along with an impractical configuration of having separate excitations for each radiator. The CRLH ME-dipole array may be seen as a practical way to achieve such radiation pattern characteristics with high gain performance.

\section{Measurement Results}

 \begin{figure*}[htbp]
\begin{center}
\subfigure[]{
\psfrag{a}[c][c][0.75]{\color{white}100~$\Omega$ Termination}
\psfrag{b}[l][c][0.75]{\color{white}50~$\Omega$ Termination}
\psfrag{c}[c][c][1]{ME-Dipole Array}
\psfrag{d}[c][c][1]{Rat-Race Coupler}
\psfrag{e}[c][c][0.75]{\color{white}(Top layer)}
\psfrag{x}[c][c][0.8]{\color{white}$x$}
\psfrag{y}[c][c][0.8]{\color{white}$y$}
\psfrag{z}[c][c][0.8]{\color{white}$z$}
\psfrag{m}[c][c][0.8]{\color{white}Via\#2}
\psfrag{n}[c][c][0.8]{\color{white}Via\#1}
\psfrag{o}[c][c][0.8]{\color{white}Via\#3}
\includegraphics[width=2\columnwidth]{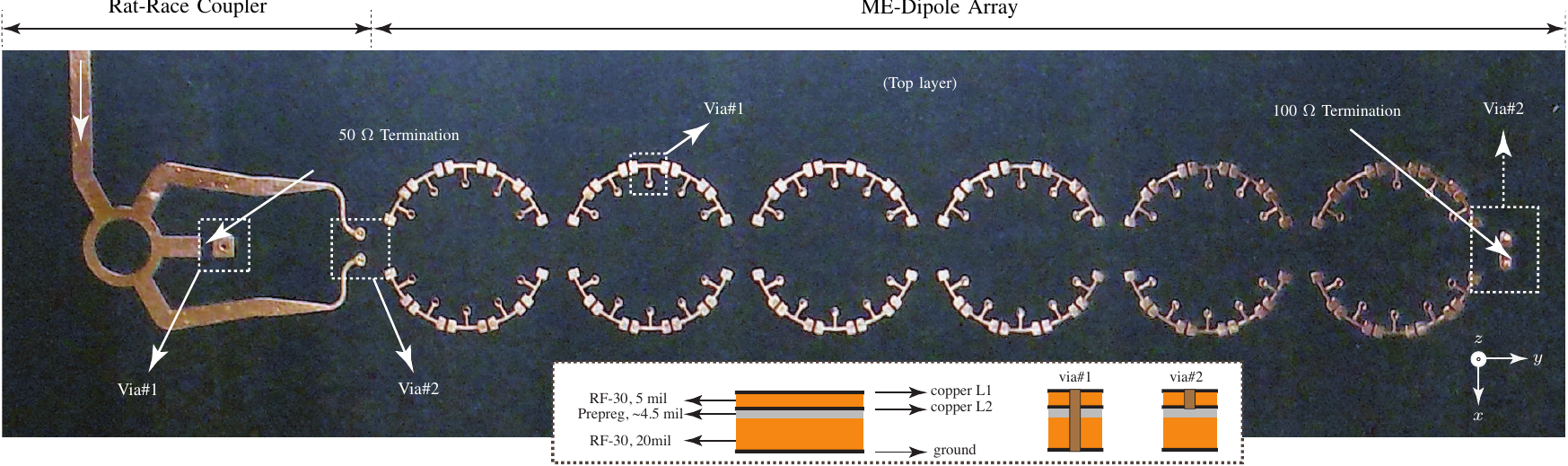}}
\subfigure[]{
\psfrag{a}[c][c][0.8]{Elevation angle $\theta$ (deg)}
\psfrag{b}[c][c][0.8]{Frequency (GHz) }
\psfrag{c}[c][c][0.8]{\shortstack{M-dipole $G_\phi(\theta)$\\ $\phi=0^\circ$ ($x-z$ plane)}}
\psfrag{d}[c][c][0.8]{\shortstack{E-dipole $G_\theta(\theta)$\\ $\phi=0^\circ$ ($x-z$ plane)}}
\psfrag{e}[c][c][0.8]{\shortstack{E-dipole (co-pol) $G_\phi(\theta)$\\ $\phi=90^\circ$ ($y-z$ plane)}}
\psfrag{f}[c][c][0.8]{\shortstack{Cross-pol $G_\theta(\theta)$\\ $\phi=90^\circ$ ($y-z$ plane)}}
\psfrag{g}[c][c][0.8]{\color{white}$f_0$}
\includegraphics[width=2\columnwidth]{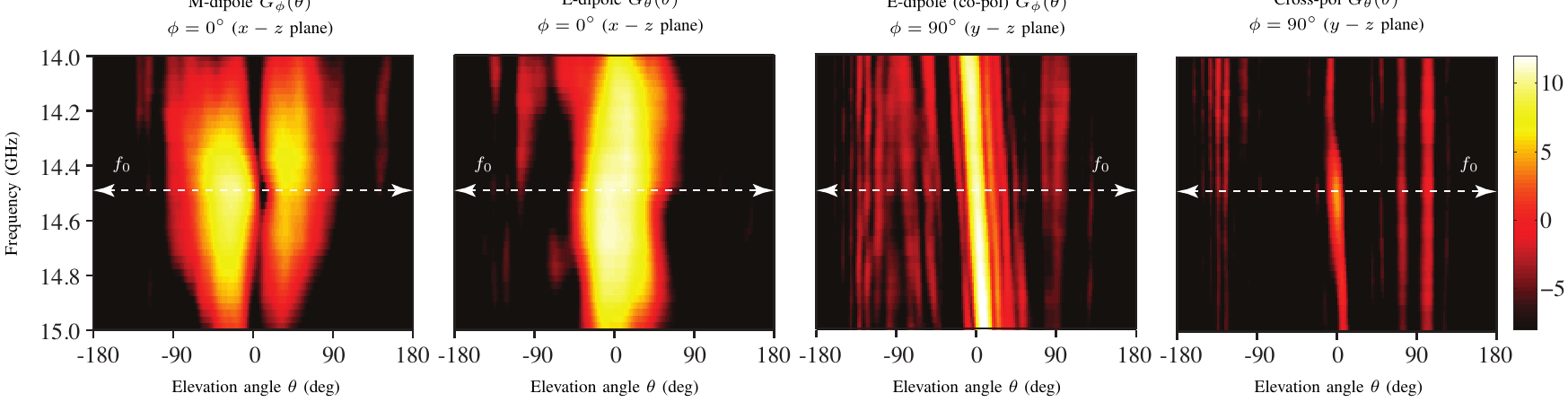}}
\subfigure[]{
\psfrag{c}[c][c][0.8]{\shortstack{M-dipole $G_\phi(\theta)$\\ $\phi=0^\circ$ ($x-z$ plane)}}
\psfrag{d}[c][c][0.8]{\shortstack{E-dipole $G_\theta(\theta)$\\ $\phi=0^\circ$ ($x-z$ plane)}}
\psfrag{e}[c][c][0.8]{\shortstack{E-dipole (co-pol) $G_\phi(\theta)$\\ $\phi=90^\circ$ ($y-z$ plane)}}
\psfrag{g}[c][c][0.8]{$f_0=14.5$~GHz}
\psfrag{f}[c][c][0.8]{$f_\text{LF}=14.25$~GHz}
\psfrag{h}[c][c][0.8]{$f_\text{RF}=14.75$~GHz}
\psfrag{x}[l][c][0.8]{Measured}
\psfrag{y}[l][c][0.8]{FEM-HFSS}
\includegraphics[width=1.3\columnwidth]{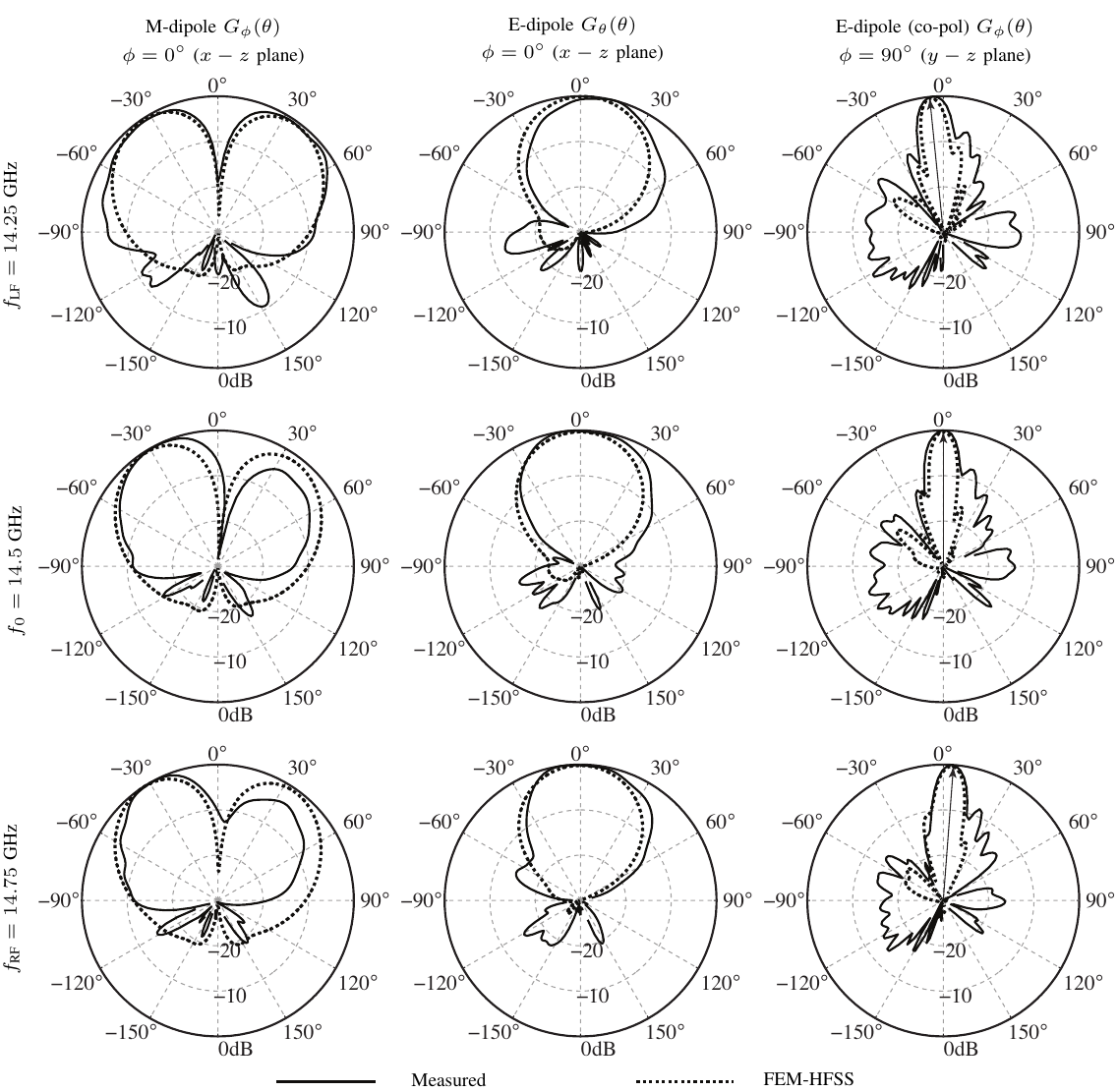}}
\subfigure[]{
\psfrag{a}[c][c][0.8]{Frequency (GHz)}
\psfrag{b}[c][c][0.8]{Measured gain, $G_\theta$ at $\phi=0^\circ$ (dB)}
\psfrag{c}[l][c][0.8]{Measured $\theta = 0^\circ$}
\psfrag{d}[l][c][0.8]{FEM-HFSS}
\psfrag{e}[c][c][0.7]{$f_0=14.5$~GHz}
\psfrag{h}[l][c][0.8]{Measured $\theta = -25^\circ$}
\psfrag{i}[l][c][0.8]{Measured $\theta = +25^\circ$}
\psfrag{f}[c][c][0.8]{Peak gain, $G_\phi$ at $\phi=0^\circ$ (dB)}
\includegraphics[width=0.65\columnwidth]{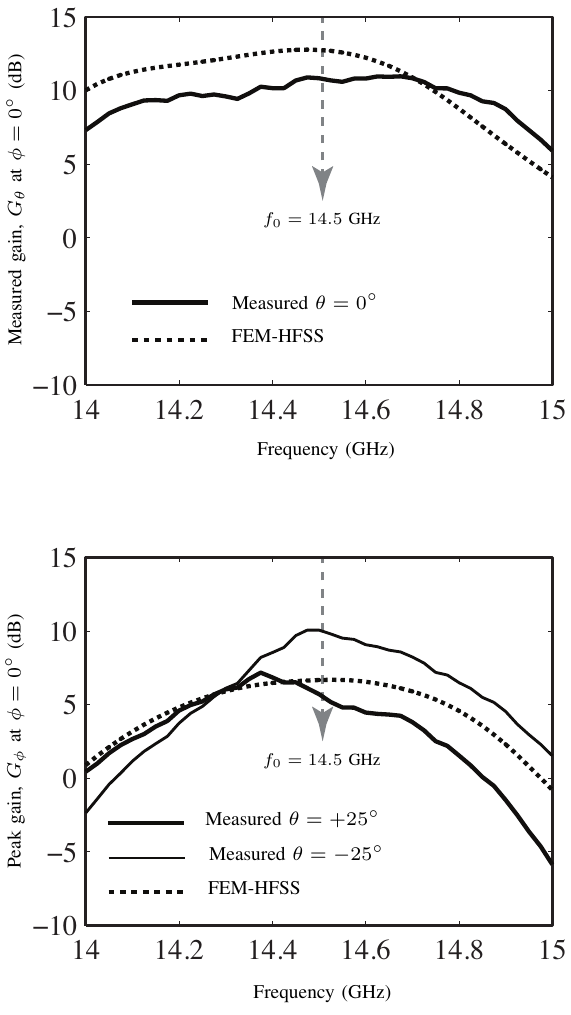}}
\caption{Fabricated CRLH TL based ME-Dipole array structure and measured radiation patterns compared with FEM-HFSS. a) Photograph of the antenna structure showing the layout patterns of the top metal layer, along with the layer and via definitions in the inset. b) Measured radiation maps of two orthogonal polarizations in $x-z$ and $y-z$ planes. c) 2D Radiation patterns of three frequencies corresponding to $f_\text{LF}=14.25$~GHz in the LH range, transition frequency $f_0=14.5$~GHz and $f_\text{RH} = 14.75$~GHz in the RH range of the CRLH structure. d) Measured peak-gain corresponding to E- and M-dipole components of the ME-dipole array. The structure consists of 2 substrates of heights, $0.508$~mm and $0.127$~mm, respectively. The structural parameters are: $p=16$~mm, $a=7$~mm, $w_s = 0.254$~mm, $w_\text{st}=0.254$~mm, $\ell_s = 1.5$~mm, $\Delta\theta_1=6.25^\circ$, $\Delta\theta_2 =16^\circ$ and $\Delta\theta_3=14^\circ$. The via radius = $0.254$~mm.} \label{Fig:Measured_Pats}
\end{center}
\end{figure*}

To experimentally demonstrate the ME-dipole antenna array, an MIM 6-ring array similar to that of Fig.~\ref{Fig:ArrayHFSS}(a) was fabricated, as shown in Fig.~\ref{Fig:Measured_Pats}(a). The differential excitation of the ME-dipole array is achieved using a 4-port rat-race coupler designed at the center frequency $f_0$, with isolated port terminated with a $50$~$\Omega$ load. The second port of the antenna array is terminated by a matched $100$~$\Omega$ load. It may thus be seen as a truncated structure where the power left at the end of the structure, is absorbed by the load. The prototype consists of two Taconic RF-30 substrates, $\varepsilon_r=3.0$, which are $0.508$~mm and $0.127$~mm thick, respectively, as shown in the inset of Fig.~\ref{Fig:Measured_Pats}(a). A prepreg layer with $\varepsilon_r=2.42$ and $\tan\delta=0.02$, from Taconic is used to connect the two substrate together, to form the multi-layer configuration. The actual thickness of the prepreg layer was not known at the time of the fabrication, and consequently, the suggested thickness of $114$~$\mu$m from the Taconic datasheet is assumed in modelling this structure in FEM-HFSS.

\begin{figure}[htbp]
\begin{center}
\psfrag{a}[c][c][1]{Frequency (GHz)}
\psfrag{b}[c][c][1]{Return loss, $|S_{11}|$ (dB)}
\psfrag{c}[l][c][0.7]{FEM-HFSS}
\psfrag{d}[l][c][0.7]{Meas.: rat-race coupler input}
\psfrag{e}[l][c][0.7]{Meas. $S^\text{diff}_{11}$ : antenna input}
\psfrag{f}[c][c][0.7]{antenna input}
\psfrag{g}[c][c][0.7]{differential probe}
\includegraphics[width=\columnwidth]{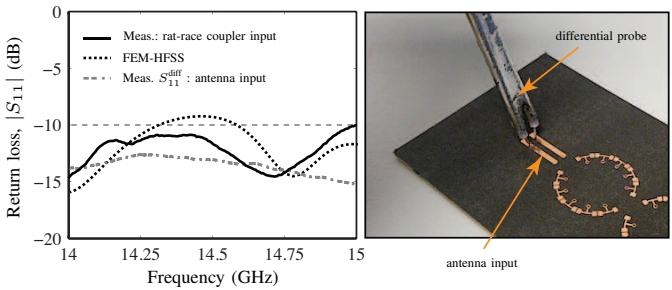}
\caption{Measured S-parameters of the ME-dipole array of Fig.~\ref{Fig:Measured_Pats}(a) at the input of the rat-race coupler and the antenna array.} \label{Fig:Measures_S}
\end{center}
\end{figure}

Figure~\ref{Fig:Measures_S} shows the measured S-parameters of the fabricated prototype at the input of the rat-race coupler. Matching with $|S_{11}|<-10$~dB is achieved across the bandwidth of interest. The matching at the input of the antenna is also shown, where a differential probe is directly used at the two input terminals of the antenna array, as shown in the photograph of Fig.~\ref{Fig:Measures_S}, to measure the input differential return loss, $S^\text{diff}_{11}$.

The antenna was next measured in an anechoic chamber, and the measured radiation maps are shown in Fig.~\ref{Fig:Measured_Pats}(b), along two principal cuts. The M-dipole response is clearly evident in the $x-z$ plane with a null at broadside, i.e. $\theta=0^\circ$, coinciding with the maximum radiation of E-dipole. Typical beam-scanning characteristics of the CRLH structure is seen in the $y-z$ plane with the beam scanning from backward ($\theta=-4^\circ$) to forward region ($\theta=+4^\circ$) including broadside radiation ($\theta=0^\circ$). This broadside radiation corresponds to $f_0=14.5$~GHz. The corresponding 2D radiation patterns are shown in Fig.~\ref{Fig:Measured_Pats}(c), for three selected frequencies, $f_\text{LF}=14.25$~GHz in the LH range, transition frequency $f_0=14.5$~GHz and $f_\text{RH} = 14.75$~GHz in the RH range of the CRLH structure. A good agreement is obtained between measurements and simulations, in all cases. The measured gain values corresponding to the E- and M-dipoles are shown in Fig.~\ref{Fig:Measured_Pats}(d), computed in the $x-z$ plane. The broadside gain of $10.84$~dB is achieved for the E-dipole component, compared to the simulated gain of 12.74~dB. Similarly the M-dipole gains are $10.06$~dB and $5.73$~dB corresponding to the two beams in ($-x$-$z$) and ($+x$-$z$) quadrants, respectively, compared to the simulated gain of 7.2 dB (two beams symmetrical in HFSS). This asymmetry can be attributed to the presence of the rat-race coupler and the SMA connector used in the measurement chamber. This can easily be corrected by choosing an connector suitable for high-frequency measurements and appropriately shielding it. The difference between the measurement and the simulated gains could be attributed to the uncertainty regarding the exact material and structural properties of the prepreg layer used, along with high sensitivity to fabrication tolerances at these frequencies.

It is to be noted that the measured transition frequency is shifted by $0.25$~GHz towards high frequencies, compared to the design frequency of $14.25$~GHz (see Fig.~\ref{Fig:ArrayHFSS}). This shift is caused by the extra prepreg layer in the fabricated prototype, which was not present in Fig.~\ref{Fig:ArrayHFSS}. Corresponding to $f_0=14.5$~GHz, the circumference of each ring in the array is about $2.1\lambda_0$ and the antenna is about $4.8\lambda_0$ long, along the y-axis.

\section{Conclusions}

A planar magneto-electric (ME) dipole antenna array based on CRLH transmission lines has been proposed and demonstrated by both full-wave and experimental results. The series elements of the CRLH transmission lines form the M-dipoles and the shunt elements form the E-dipoles. The infinite-wavelength propagation characteristic of CRLH transmission lines has been exploited so as to form high-gain magnetic radiators in conjunction with conventional electric radiators which are excited simultaneously using a single differential feed. The travelling-wave type nature of the proposed structure enables the formation of directive arrays with high-gain characteristics. 

By operating the antenna at the transition frequency of the antenna, a uniform current can be maintained across the rings to achieve a high peak M-dipole gain. However, at frequencies different from the transition frequency, the gain drops from its peak values as a result of formation of non-uniform currents, thereby making the proposed structure narrowband in nature. In addition, due to the CRLH balancing requirements, the series and shunt elements cannot be tuned to control the E- and M-dipole gains independently. However, by controlling the size of the ring, the E- and M-dipole gains may be equalized. 

The proposed ME-dipole antenna array is a dual-polarized antenna. While offering an unconventional radiation characteristic combining magnetic and electric dipole radiation, it may be customized and adapted to different requirements, such as in applications involving beam-switching or space diversity. For example, by simultaneously exciting the two differential ports of the structure with an appropriate polarity, using switches and hybrids, one may selectively excite the magnetic and electric dipole component only \cite{Sam_PD_CRLHDIFF}. Alternatively, the proposed ME-dipole array antenna may be operated in a resonant configuration, using a short or a open circuit at the second port of the array, to selectively excite either the series or shunt resonances of the CRLH structure, which are responsible for the M- and E-dipole radiation \cite{Caloz-MTM-Book, DualPolCRLH3}. The antenna, depending on the scheme used, may thus be used as a purely magnetic loop array, or an electric-dipole array, or in a hybrid configuration where the two radiations are used in tandem in a time-domain multiplexing manner. The proposed antenna thus exhibits great flexibility and versatility in its radiation characteristics, and may lead to innovative phased array systems and applications in the future.

\section*{Acknowledgement}
Authors would like to thank Mr. Zilong Ma at the University of Hong Kong and Dr. Kwok Kan So at the City University of Hong Kong, for their help in fabrication and measurements of various prototypes. Authors would like to particularly thank the State Key Laboratory of Millimeter Waves at the City University of Hong Kong for providing the measurement facilities. This work was supported in part by HK ITP/026/11LP, HK GRF 711511, HK GRF 713011, HK GRF 712612, and NSFC 61271158.

\bibliographystyle{IEEEtran}
\bibliography{References_ME}

\begin{thebibliography}{10}
\providecommand{\url}[1]{#1}
\csname url@samestyle\endcsname
\providecommand{\newblock}{\relax}
\providecommand{\bibinfo}[2]{#2}
\providecommand{\BIBentrySTDinterwordspacing}{\spaceskip=0pt\relax}
\providecommand{\BIBentryALTinterwordstretchfactor}{4}
\providecommand{\BIBentryALTinterwordspacing}{\spaceskip=\fontdimen2\font plus
\BIBentryALTinterwordstretchfactor\fontdimen3\font minus
  \fontdimen4\font\relax}
\providecommand{\BIBforeignlanguage}[2]{{%
\expandafter\ifx\csname l@#1\endcsname\relax
\typeout{** WARNING: IEEEtran.bst: No hyphenation pattern has been}%
\typeout{** loaded for the language `#1'. Using the pattern for}%
\typeout{** the default language instead.}%
\else
\language=\csname l@#1\endcsname
\fi
#2}}
\providecommand{\BIBdecl}{\relax}
\BIBdecl

\bibitem{ME_Review}
K.~M. Luk and B.~Wu, ``The magnetoelectric dipole, a wideband antenna for base
  stations in mobile communications,'' \emph{Proceeding of the IEEE}, vol. 100,
  no.~7, pp. 2297--2307, Jul. 2012.

\bibitem{LoopArray_Vehicular}
T.~Nakanishi, T.~Yoshida, A.~Ishida, H.~Uno, and Y.~Saito, ``Multiple-loop
  array antenna with switched beam for short-range radars,'' \emph{Vehicular
  Technology Conference, 2006. VTC-2006 Fall. 2006 IEEE 64th}, pp. 1--5, Sept.
  2006.

\bibitem{EM_MIMO}
J.~Xiong, M.~Zhao, H.~Li, Z.~Ying, and B.~Wang, ``Collocated electric and
  magnetic dipoles with extremely low correlation as a reference antenna for
  polarization diversity {MIMO} applications,'' \emph{IEEE Antennas Wirel.
  Propagat. Lett.}, vol.~11, no.~6, pp. 423--426, Apr. 2012.

\bibitem{Sam_PD_CRLHDIFF}
S.~Abielmona, H.~V. Nguyen, and C.~Caloz, ``{CRLH LWA} with polarization
  diversity using equalized common and differential modes,'' \emph{in IEEE AP-
  S Int. Antennas Propagat. (APS), Chicago, IL}, Jul. 2012.

\bibitem{Loop_Array}
K.~Wei, Z.~Zhang, Z.~Feng, and M.~F. Iskander, ``A {MNG-TL} loop antenna array
  with horizontally polarized omnidirectional patterns,'' \emph{IEEE Trans.
  Antennas Propagat.}, vol.~60, no.~6, pp. 2702--2709, Jun. 2012.

\bibitem{Loop_Array_old}
M.~Callendar, ``Aperiodic loop antenna arrays,'' \emph{Antennas and Propagation
  Society International Symposium, 1969}, pp. 193--198, Dec. 1969.

\bibitem{PhasedArray}
P.~Loghmannia, M.~Kamyab, M.~N. Ranjbar, and R.~Rezaiesarlak, ``Miniaturized
  low-cost phased array antenna using siw slot elements,'' \emph{IEEE Microw.
  Wireless Compon. Lett.}, pp. 1434 -- 1437, Nov. 2012.

\bibitem{Equal_E_H}
A.~Clavin, ``A new antenna feed having equal {E}-and {H}-plane patterns,''
  \emph{IRE Trans. Antennas Propog.}, vol. AP-2, pp. 113--119, Apr. 1954.

\bibitem{Slot_Dipole}
R.~W.~P. King and G.~H. Owyang, ``The slot antenna with coupled dipoles,''
  \emph{IRE Trans. Antennas Propog.}, vol. AP-8, pp. 136--143, Mar. 1960.

\bibitem{Colocated}
P.~L. Overfelt, D.~R. Bowling, and D.~J. White, ``A colocated magnetic loop,
  electric dipole array antenna (preliminary results),'' \emph{NAWCWPNS Tech
  Pub 8212, Naval Air Warfare Center Weapons Division, China Lake, CA}, Sept.
  1994.

\bibitem{Luk_EM}
M.~Li and K.-M. Luk, ``A differential-fed magneto-electric dipole antenna for
  {UWB} applications,'' \emph{IEEE Trans. Antennas Propagat.}, vol.~61, no.~1,
  pp. 92--99, Jan. 2013.

\bibitem{Luk_ME}
K.~M. Luk and H.~Wong, ``A new wideband unidirectional antenna element,''
  \emph{IJMOT}, vol.~1, pp. 35--44, Apr. 2006.

\bibitem{CRLH_Mono}
C.~Caloz, T.~Itoh, and A.~Rennings, ``{CRLH} metamaterial leaky-wave and
  resonant antennas,'' \emph{IEEE Antennas Propagat. Mag.}, vol.~50, no.~5, pp.
  25--39, Oct. 2008.

\bibitem{Caloz-MTM-Book}
C.~Caloz and T.~Itoh, ``Electromagnetic metamaterials, transmission line theory
  and microwave applications,'' \emph{Wiley - IEEE Press}, 2005.

\bibitem{DualPolCRLH1}
M.~R.~M. Hashemi and T.~Itoh, ``Dual-mode leaky-wave excitation in symmetric
  composite right/left-handed structure with center vias,'' \emph{in Proc. IEEE
  MTT-S Int. Microw. Symp., Anaheim, CA}, vol. 100, pp. 9--12, May. 2010.

\bibitem{DualPolCRLH2}
------, ``Coupled composite right/left-handed leaky-wave transmission-lines
  based on common/ differential-mode analysis,'' \emph{IEEE Trans. Microw.
  Theory Tech.}, vol.~58, no.~12, pp. 3645--3656, Dec. 2010.

\bibitem{Balanis-Antenna-Book}
C.~A. Balanis, ``Advanced engineering electromagnetics,'' \emph{John Wiley and
  Sons}, 1989.

\bibitem{MTM_Eleft_Book}
G.~Eleftheriades and K.~Balmain, ``Negative-refraction metamaterials:
  Fundamental principles and applications,'' \emph{John Wiley \& Sons and IEEE
  Press}, 2005.

\bibitem{Capolino_Eleft_Book}
F.~Capolino, ``Metamaterials handbook: Applications of metamaterials,''
  \emph{CRC Press}, 2009.

\bibitem{Caloz_MT_2009}
C.~Caloz, ``Perspectives on \text{EM} metamaterials,'' \emph{Materials Today},
  vol.~12, no.~3, pp. 12--30, March 2009.

\bibitem{MTM2}
M.~A.~Y. Abdalla, K.~Phang, and G.~V. Eleftheriades, ``A planar electronically
  steerable patch array using tunable pri/nri phase shifters,'' \emph{IEEE
  Trans. Microw. Theory Tech.}, vol.~57, no.~3, pp. 531--541, Mar. 2009.

\bibitem{MTM3}
Y.-K. Jung and B.~Lee, ``Beam scannable patch array antenna employing tunable
  metamaterial phase shifter,'' \emph{IEEE Antennas and Propagation Society
  International Symposium (APSURSI)}, vol.~57, no.~3, pp. 1--2, Jul. 2012.

\bibitem{CRLH_Loop}
A.~Locatelli, A.~Capobianco, S.~Boscolo, D.~Modotto, M.~Midrio, and C.~D.
  Angelis, ``Low-profile {CRLH} omnidirectional loop antenna for mobile
  wireless communications,'' \emph{Microwave Conference (EuMC), 2012 42nd
  European}, pp. 401--403, Oct. 2012.

\bibitem{Zeroth_order_CRLH_Loop}
J.-G. Lee and J.-H. Lee, ``Zeroth order resonance loop antenna,'' \emph{IEEE
  Trans. Antennas Propagat.}, vol.~55, no.~3, pp. 994--997, Mar. 2007.

\bibitem{Leaky_Book}
C.~Caloz, D.~R. Jackson, and T.~Itoh, ``Frontiers in antennas,'' \emph{F.
  Gross, Ed. New York: McGraw-Hill}, 2010.

\bibitem{Leaky_jackson}
D.~R. Jackson, C.~Caloz, and T.~Itoh, ``Leaky-wave antennas,'' \emph{Proc.
  IEEE}, vol. 100, no.~7, pp. 2194--2206, Jul. 2012.

\bibitem{Otto_Leaky}
S.~Otto, A.~Rennings, K.~Solbach, and C.~Caloz, ``Transmission line modeling
  and asymptotic formulas for periodic leaky-wave antennas scanning through
  broadside,'' \emph{IEEE Trans. Antennas Propagat.}, vol.~59, no.~10, pp.
  3695--3709, Oct. 2010.

\bibitem{Diff_yang}
N.~Yang, C.~Caloz, and K.~Wu, ``Greater than the sum of its parts,''
  \emph{Microw. Mag.}, vol.~11, no.~4, pp. 69--82, Jun. 2010.

\bibitem{Diff_Amps}
J.~Karki, ``Fully differential amplifiers,'' \emph{Texas Instruments, Texas,TI
  Application Rep. SLOA054D}, Jan. 2002.

\bibitem{CRLH_Active}
F.~P. Casares-Miranda, C.~Camacho-Pe–alosa, and C.~Caloz, ``High-gain active
  composite right/left-handed leaky-wave antenna,'' \emph{IEEE Trans. Antennas
  Propagat.}, vol.~54, no.~8, pp. 2292--2300, Aug. 2006.

\bibitem{CRLH_Distributed_Amplifier}
K.~Mori and T.~Itoh, ``Distributed amplifier with {CRLH}-transmission line
  leaky wave antenna,'' \emph{Proceedings of the 38th European Microwave
  Conference}, pp. 686--689, Oct. 2008.

\bibitem{Caloz_AWPL}
C.~Caloz and T.~Itoh, ``Array factor approach of leaky-wave antennas and
  application to 1{D}/2{D} composite right/left-handed ({CRLH}) structures,''
  \emph{IEEE Microw. Wireless Compon. Lett.}, vol.~14, no.~6, pp. 274--276,
  Jun. 2004.

\bibitem{DualPolCRLH3}
M.~R.~M. Hashemi and T.~Itoh, ``Evolution of composite right/left-handed
  leaky-wave antennas,'' \emph{Proceedings of the IEEE}, vol.~99, no.~10, pp.
  1746 -- 1754, Oct. 2011.

\end{thebibliography}

\end{document}